\documentclass{article}

\usepackage[final]{neurips_2021_ml4ps}

\usepackage[utf8]{inputenc} 
\usepackage[T1]{fontenc}    
\usepackage{hyperref} 
\usepackage{comment}
\usepackage{url}            
\usepackage{booktabs}       
\usepackage{amsfonts}       
\usepackage{nicefrac}       
\usepackage{microtype}      
\usepackage{xcolor}         
\usepackage{graphicx}
\usepackage{amsmath}
\title{Probabilistic segmentation of overlapping galaxies for large cosmological surveys.}

\newcommand{\unet}{U-Net~}
\newcommand{\unetend}{U-Net}
\newcommand{\punet}{PUnet~}
\newcommand{\punetend}{PUnet}

\author{
  Hubert Bretonnière~\thanks{hubert.bretonniere@universite-paris-saclay.fr} \\
  Institut d'Astrophysique Spatiale\\
  Université Paris-Saclay\\
  France
  \AND
  Alexandre Boucaud \\
  Laboratoire Astroparticule et Cosmologie\\
  Université de Paris\\
  France\\
  \AND
  Marc Huertas-Company\\
  Departamento de Astrofisicia\\
  Universidad de La Laguna\\
  Spain
  
}

\begin{document}

\maketitle

\begin{abstract}
   Encoder-Decoder networks such as U-Nets have been applied successfully in a wide range of computer vision tasks, especially for image segmentation of different flavours across different fields.  Nevertheless, most applications lack of a satisfying quantification of the uncertainty of the prediction. Yet, a well calibrated segmentation uncertainty can be a key element for scientific applications such as precision cosmology. In this on-going work, we explore the use of the probabilistic version of the \unetend, recently proposed by \citet{PUnet}, and adapt it to automate the segmentation of galaxies for large photometric surveys. We focus especially on the probabilistic segmentation of \emph{overlapping} galaxies, also known as blending. We show that, even when training with a single ground truth per input sample, the model manages to properly capture a pixel-wise uncertainty on the segmentation map. Such uncertainty can then be propagated further down the analysis of the galaxy properties. To our knowledge, this is the first time such an experiment is applied for galaxy deblending in astrophysics.
\end{abstract}

\section{Introduction}
The new generation of extragalactic imaging surveys of galaxies, such as Euclid \citep{euclid_survey} or LSST \citep{lsst}, will deliver an unprecedented volume of data. These new surveys also imply more demanding scientific requirements which are translated into new challenges from the data analysis perspective. The deeper the survey, the larger the density of detected sources, leading to  an increase of the probability of observing overlapping galaxies due to projection effects. We call those objects \emph{blended}, and the task of identifying and treating that confusion, \emph{deblending}. Misidentifying blended galaxies significantly impacts the error budget and can even prevent reaching the scientific requirements for precision cosmology \citep{blending_lsst}. 

Accurately identifying blended galaxies is therefore a key task to guarantee the full potential of the forthcoming cosmological surveys. Several machine learning (ML) based \citep{boucaud, arcelin} and non-ML based solutions \citep{scarlett} have been proposed, although the issue remains open. In particular, a robust
quantification of uncertainty to be propagated into the final error budget generally lacks from the current approaches.

In this work in progress,  we adapt the Probabilistic U-Net (hereafter \punetend) architecture proposed by \citet{PUnet} to work on galaxy images. We namely modify the loss function to deal with very unbalanced datasets and demonstrate that the  Probabilistic Deep Learning model provides meaningful segmentation uncertainties while obtaining
competitive scores both in completeness and purity, even when trained from a unique ground truth.

\begin{figure}[!htb]
\centering
\includegraphics[width=\linewidth]{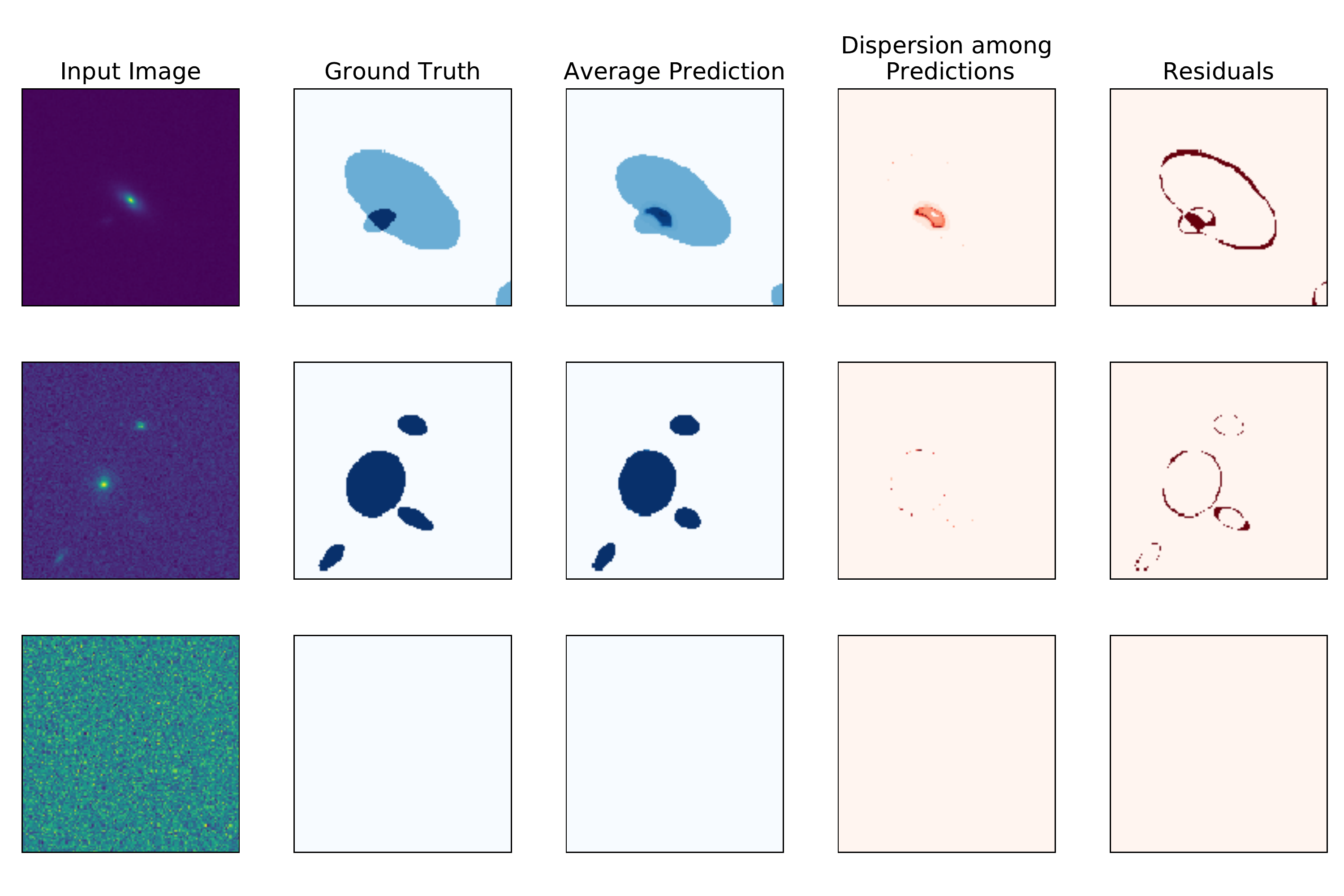}
        \caption{Examples of outputs of the probabilistic segmentation model on our test dataset. Each row shows a different case. Columns from left to right: input image; ground truth; average of $50$ realisations; standard deviation of the realisations; pixel-wise absolute difference between truth and prediction. The first row highlights the informative uncertainty for the segmentation with overlapping galaxies, along with the capacity to correctly predict truncated objects. The second row shows that the uncertainty follows the residuals at the border of the galaxy, where the predicted segmentation is not well defined. The last row acts as a null case by showing the capacity of our model to identify a pure background image and to not over-segment.}
        \label{fig:predictions}
\end{figure}

\section{Model}

The global architecture of our model follows the one used by \citet{PUnet}. The architecture of the \punet is made of two parts. On the one hand, there is a classical \unet \citep{Unet} architecture, whose goal is to create realistic segmentation maps from the input galaxy images. This constitutes the deterministic part of the model. On the other hand, we have an encoder that compresses the images into a Gaussian latent space. This latent space is sampled and concatenated to the output of the \unet before a final convolution layer followed by a  softmax activation to produce a non-deterministic output. The output is then thresholded to produce the desired segmentation map with overlapping regions.

During training, the latent space is regularised using a second encoder that takes as input the ground truth segmentation in addition to the input image. At inference time, that second encoder is removed and the samples are taken only from the latent space of the first encoder. That sampling allows one to have multiple  realisations of the segmentation and hence an estimate of the uncertainty at the pixel level. See appendix Section \ref{sec:architecture} and Figure \ref{fig:model} for more details about the specific architecture used.

The loss function is a weighted sum of two terms: a reconstruction loss and a Kullback--Leibler Divergence \citep{KL} between the Gaussian distributions of the two latent space encoders. The former takes care of producing realistic segmentation maps while the latter ensures that the inferred distribution without the ground truth preserves a memory of the ground truth.  

We introduce two main modifications to the original \punet model. First, we replace the reconstruction loss by a Dice loss \citep{dice}, more suited to our unbalanced segmentation problem (see Section \ref{sec:data}). We weight the Dice loss so that a wrong prediction on a blend as more importance than a wrong prediction on the background, which dominates the image. Second, we set our model to train from a single ground truth. While the original \punet model was used to cope with \emph{ambiguous} segmentations, i.e images with different possible ground truths according to experts, in our case, we only have one ground truth. However, we show that the model can  still capture an informative uncertainties from the diversity of the dataset.

\section{Data}\label{sec:data}

Our data consists of $128\times\,128$ image stamps which have been extracted from a large simulated galaxy field, meant to reproduce the typical properties of future space missions both in terms of signal-to-noise and spatial resolution. In this first proof-of-concept work, the galaxies in the field are simulated as pure analytic Sersic profiles \citep{sersic}, using the \texttt{Galsim} software \citep{galsim}. A noise realisation is then added to the stamp. The true segmentation maps are created by taking all the pixels of the galaxy that are above the noise level of the image. It produces ternary maps, where all pixels belonging to the background are set to $0$ and the pixels of the galaxy are set to $1$ and overlapping regions to $2$. We simulate $4$ galaxy fields of $25\,000\times\,25\,000$ pixels, and extract the stamps from them, which represents a training set of $~150\,000$ images. We preprocess the images with an inverse hyperbolic sine (\texttt{arcsinh}), and then a normalisation by the maximum of each stamp. 

 We emphasise three properties of the dataset which contribute to make the problem more complex. First, the three classes are strongly imbalanced. $\sim96\%$ of the pixels belong to the background, $\sim3\%$ to isolated galaxy, and $<1\%$ correspond to the overlapping regions we would like to focus on. The imbalance between the classes is resolved with the weighted Dice loss, we do not apply any data augmentation. Second, there is a significant fraction of images with no galaxies on them, which we call background stamps. The model could easily fall in a mode where it predicts only background. Finally, because we arbitrarily cut our entire fields into stamps following a fixed grid, a significant amount of galaxies can be cut in the border of the stamps. It means that our algorithm needs to learn the shape of a galaxy, but also recognise a truncated galaxy profile. 
 
 Our test sets is done following the exact same procedure, and contains $\sim 3000$ isolated galaxies and the same number of blended objects.

\section{Results}\label{sec: results}

\subsection{Images}\label{sec:images}
We  illustrate in Figure \ref{fig:predictions} a set of typical prediction cases extracted from the test dataset to provide a qualitative overview of the model behaviour. Three interesting properties can be highlighted from the Figure. 1 - The capacity of the model for segmenting overlapping galaxies while retaining some uncertainty in the prediction on the overlapping region; 2 - The accurate segmentation of multiple galaxies in the field, including truncated objects, with residuals being concentrated towards the galaxy outskirts and 3 - The ability of the network to deal with empty stamps. That last case acts as a null test for the network and represents an improvement with respect to the model from \citet{boucaud} which would fail at predicting only a background noise.

\begin{table}
\centering
\begin{tabular}{l|c|c}
                    & Completeness & Purity  \\ 
        \hline
        Isolated    & 99.1       &   98.5   \\ 
        \hline
        Blended     & 87.3       & 93.6  
        \end{tabular}
        \caption{Global scores on a large field of galaxies.}
        \label{tab:c/p}
\end{table}

\subsection{Completeness, Purity and IoU} 
We use three standard metrics to evaluate the results: the completeness, the purity, and the classical segmentation metric: Intersection over Union (IoU), between the ground truth and the prediction.

\begin{figure}[!htb]
    \centering
    \begin{minipage}{.45\linewidth}
    \centering
    \includegraphics[width=\linewidth, height=0.2\textheight]{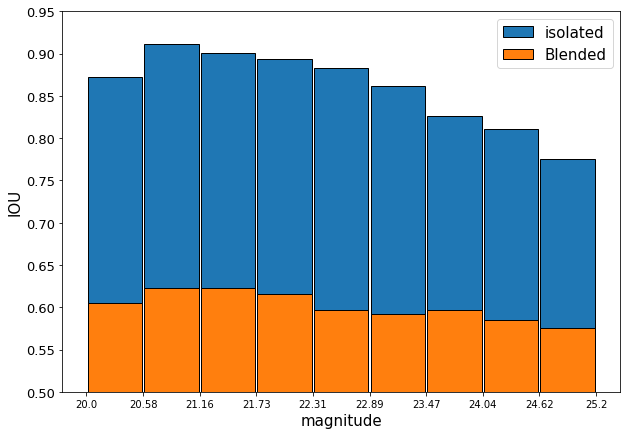}
    \caption{Intersection over Union in bins of magnitude, for isolated and blended galaxies. Larger magnitudes mean fainter objects with lower signal-to-noise.}
    \label{fig:IOU}
        
    \end{minipage}%
    \hspace{1cm}
    \begin{minipage}{0.45\linewidth}
        \centering
        \includegraphics[width=\linewidth, height=0.2\textheight]{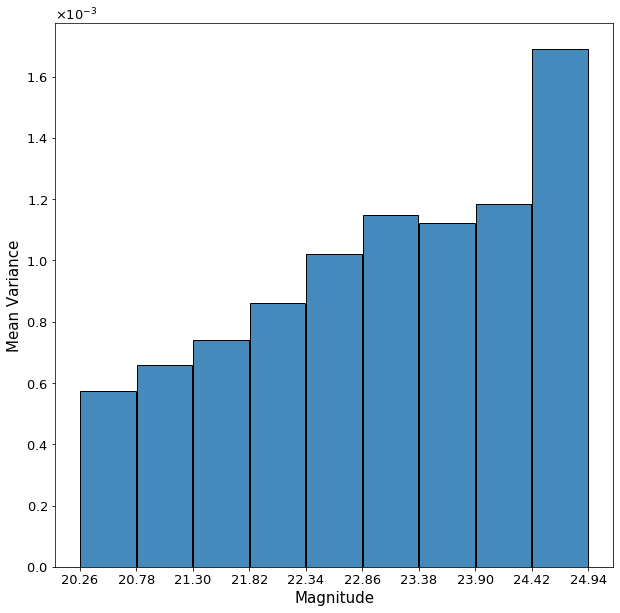}
        \caption{Mean variance of the detected object by bin of magnitude. The increase of the variance with magnitude is an expected behavior for a physically motivated uncertainty.}
    \label{fig:var_mag}
    \end{minipage}
\end{figure}

The results are summarised in Table \ref{tab:c/p} and Figure \ref{fig:IOU}. Our model is able to detect almost all isolated galaxies up to magnitude $25.2$, the limiting magnitude of our simulated galaxies. The purity is also very high, with a value of $98.5$. For the blended galaxies, the model also reaches high scores as compared to other existing approaches. We measure a completeness of $87$ and a purity of $93.6$. Our model reaches an IoU above $0.8$ at any magnitude, and up to $0.9$ for bright objects (small magnitude). The results follow the same trend for blended objects, with a lower value (as expected), and still above $0.55$ for any magnitude.

\subsection{Uncertainty calibration}
The third column of the Figure \ref{fig:predictions} shows the standard deviation of $50$ different predictions for the same input. We appreciate a sensitive behavior of the variance.
We see a larger variance at the border of the galaxies, where the flux is low, which means a harder prediction, while it is low at the centre of the object, even for fainter galaxies. We also show in Figure \ref{fig:var_mag} that the variance of objects increases with magnitude, which should be expected if the uncertainty is physically meaningful. 

\section{Future developments}
This work is an ongoing project. Based on the promising preliminary results we envision several improvements.

 We plan to make the weights of the dice losses trainable and fine tune the other hyper parameters. We will compare to state of the art classical detection algorithms \citep[e.g.][]{sextractor}. For now, we have limited our analysis to only one data set with similar properties than the training set. We will further investigate the behaviour of our model on more realistic data (observations or complex simulations), with other noise levels, etc. We will pursue the calibration of the uncertainty to better understand the link between the variance of the prediction and the aleatoric uncertainty. Finally,  we may use a sliding window for the prediction of big fields instead of a fix cut grid (see appendix~\ref{app:big_field}).

\section{Conclusion}
This work shows the first application of a deep learning probabilistic model to the segmentation of astronomical images. The model has been modified to take into account the specific properties of those images such as a low signal-to-noise and strong unbalancing. Our preliminary results indicate that our method achieves competitive performance for the detection of isolated and blended galaxies in large photometric fields, while providing an estimate of the model uncertainty. 

\bibliographystyle{aa.bst}
\bibliography{biblio}

\newpage

\appendix

\section{Architecture and Training}\label{sec:architecture}
\begin{figure}[!htb]
\centering
\includegraphics[width=\linewidth]{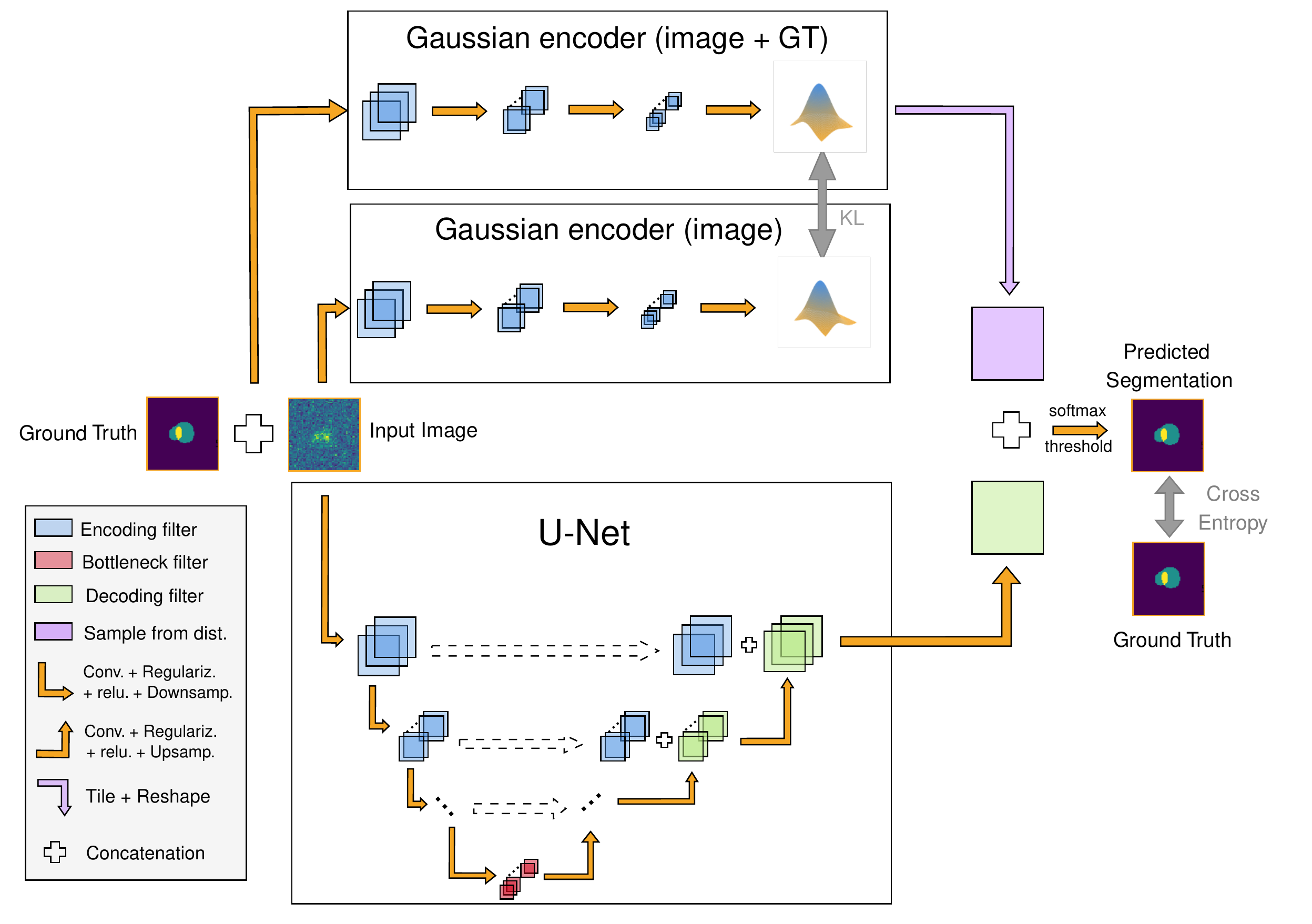}
        \caption{Illustration of the Probabilistic \unet model used in this work. The encoder at the top which takes the ground truth as input is only used during training to regularise the Gaussian latent space of the second encoder.}
        \label{fig:model}
\end{figure}

Our architecture, whose schematic is shown on Figure \ref{fig:model}, is adapted from \citet{PUnet}. The precise shape of the encoder is as follows: the filter size of the convolution layers is  $[32, 64, 128, 192, 192, 192, 256]$. Each layer uses a $3\times3$ kernel and is followed by a \emph{ReLU} activation and an average pooling layer of a factor 2.

For the \unetend, the decoder follows a symmetrical architecture from the encoder described above, with bi-linear interpolation to increase the dimensions and skip connections between the encoder and the decoder.

For the probabilistic encoder, we use the architecture from above and the final layer is flattened into a vector that encodes a $10$-dimensional Gaussian distribution which is the latent space.

 The decoder of the \unet follows a symmetrical architecture, with a bi-linear interpolation to increase the dimensions. The final CNN which takes the sample from the Gaussian and the output of the \unet is made of two layers with one filter and a $1\times1$ kernel.
 
We train our model with $150\,000$ images of $128\times128$ pixels, and a batch size of 32 and a learning rate of $10^{-3}$. To counter the imbalanced classes, we put a weight of $0.9$ (resp. $0.5$ and $0.4$) on the reconstruction of the blended regions (resp. the isolated regions and background). The model converges after $5$ epochs, during which the learning rate follows an exponential decay to $3.10^{-4}$. The training requires ~$2$ hours on a NVIDIA Tesla P40 GPU, hosted by the Paris Observatory.

\section{Segmentation on a large galaxy field}
\label{app:big_field}

\begin{figure}
    \centering
    \includegraphics[scale=0.3]{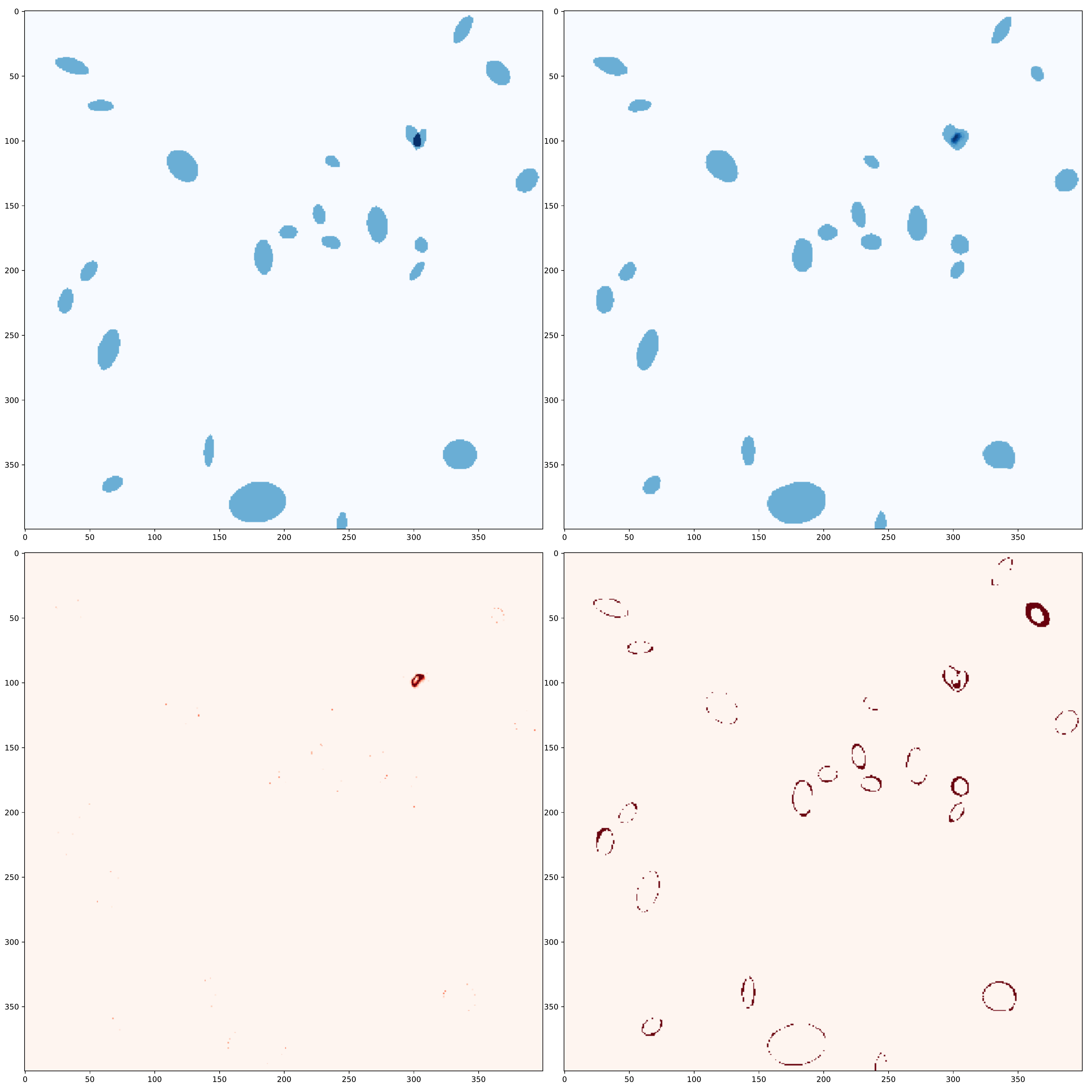}
    \caption{Example of the capacity of our model on a larger field. The field is reconstructed from stamps as described Section \ref{sec: results}. From left to right and top to bottom: the ground truth, the average prediction, the dispersion among the predictons, and the residuals.}
    \label{fig:residual}
\end{figure}

The final goal of our algorithm is to work on large galaxy fields. To investigate our results in such a scenario, we use a two-step detection. First we segment the images with our model and use that output as a detection step. Then, we use this first detection to create a new test set with stamps centred on each detected object. Doing so, we obtain a score for each galaxy. We then use the \texttt{scikit-image} library \citep{scikit-image} to clean the detected stamp and only keep the centred galaxy. Then we can reconstruct the predicted large field by summing the individual segmentation maps. 
To compare with the original field, we create residual maps from the difference with the reconstructed field. A typical residual map can be seen on Figure \ref{fig:residual}.

\end{document}